\documentclass[%
 reprint,
superscriptaddress,
 amsmath,amssymb,
 aps,
]{revtex4-2}

\usepackage{graphicx}
\usepackage{dcolumn}
\usepackage{bm}
\usepackage[mathlines]{lineno}

\begin{document}

\preprint{APS/123-QED}

\title{Generation of synchronized high-intensity x-rays and mid-infrared pulses by Doppler-shifting of relativistically intense radiation from near-critical-density plasmas}

\author{Nikita A. Mikheytsev}
 \email{welt992@gmail.com}
 \affiliation{Lobachevsky State University of Nizhny Novgorod,
 603022 Nizhny Novgorod, Russia}
\author{Artem V. Korzhimanov}
 \email{artem.korzhimanov@ipfran.ru}
 \affiliation{Lobachevsky State University of Nizhny Novgorod,
 603022 Nizhny Novgorod, Russia}
 \affiliation{Federal Research Center Institute of Applied Physics of the Russian Academy of Sciences,
 603950 Nizhny Novgorod, Russia}

\date{\today}

\begin{abstract}
It is shown that when relativistically intense ultrashort laser pulses are reflected from the boundary of a plasma with a near-critical density, Doppler frequency shift leads to generation of intense radiation both in the high-frequency, up to the X-ray, range, and in the low-frequency, mid-infrared, range. The efficiency of energy conversion into the wavelength range of greater than 3 $\mu$m can reach several percent, which makes it possible to obtain relativistically intense pulses in the mid-infrared range. These pulses are synchronized with high harmonics in the ultraviolet and X-ray ranges, which opens up opportunities for high-precision pump-probe measurements, in particular, laser-induced electron diffraction and transient absorption spectroscopy.
\end{abstract}

\maketitle

\paragraph{Introduction}

The development of methods for generating and amplifying femtosecond laser pulses has led to the possibility of obtaining pulses with a peak power of up to several petawatts \cite{Danson2015b, Sung2017, Lureau2020, Radier2022} and intensities above 10$^{23}$~W/cm$^2 $ \cite{Yoon2019, Yoon2021}. The interaction of such radiation with matter leads to efficient generation of charged particle beams \cite{Esarey2009, Macchi2013} and radiation of various ranges \cite{Corde2013, Albert2016}. In particular, much attention has been paid to the problem of generating high harmonics in the range of up to several keV from the surface of solid-density targets \cite{Dromey2007, Teubner2009}, whose efficiency can reach tens of percent. In a number of papers, possible mechanisms for generating low-frequency radiation in the terahertz \cite{Yoshii1997, Leemans2003, Gopal2013, Liao2015, Liao2016, Kwon2018, Dechard2018, Herzer2018, Dechard2019, Zeng2020a} and mid-infrared ranges \cite{Nie2018, Zhu2019a, Kulagin2019, Nie2020a, Golovanov2021, Kulagin2021, Siminos2021} has been also discussed, however, in general, this direction is still poorly understood.

Similar to the terahertz, the mid-infrared range, the lower limit of which we will consider equal to 3 $\mu$m, is poorly mastered due to the fact that the radiation frequency here is too low for efficient laser generation, but too high for electronic generation methods. At the same time, in recent years there has been a sharp increase in the efficiency of sources in this range based on the parametric amplification of light. Peak pulse powers have now been reached at the subterawatt level in the range of 3--4 $\mu$m \cite{Mitrofanov2015} and the gigawatt level in the range of 5--9 $\mu$m \cite{Wilson2019}, which made it possible to achieve relativistic intensities and, in particular, to demonstrate the generation of relativistic high harmonics \cite{Mitrofanov2020, Mitrofanov2020a, Mitrofanov2021}.

In spite of their lack, mid-infrared sources are extremely in demand in a large number of spectrographic applications \cite{Meshcherinov2020}. In particular, it is of interest to use them for laser-induced electron diffraction, in which the electronic configuration of a molecule is determined by the diffraction signal from an electron of the same molecule, ionized by a powerful mid-IR pulse and scattered by the molecule during reverse motion \cite{Pullen2015}. In pump-probe experiments with controlled delay, this makes it possible to study the dynamics of the electronic configuration with femtosecond time resolution. As a pump, among other things, ultraviolet or X-ray radiation can be used, which produces either ionization of the molecule, or its excitation. Subfemtosecond synchronization of low-frequency and high-frequency radiation is also needed in the field of transient absorption spectroscopy, in which an object is excited by a mid-infrared pulse, and short-duration ultraviolet radiation is used as a probe \cite{Schultze2013, Chang2021}. Thus, synchronized sources of high-frequency and low-frequency radiation of high intensity are in demand.

In this work, we pay attention to the fact that, along with the generation of high harmonics, when intense laser radiation is reflected from the plasma surface, generation of low-frequency radiation is also observed. The mechanism of its generation is, in fact, the same: the Doppler frequency shift. However, unlike high harmonics, which are observed when the boundary moves towards the laser radiation, the low-frequency pulses are observed when it moves towards the plasma. Previously, this effect was not emphasized, apparently due to the fact that the efficiency of such generation in most cases is low due to the low velocity of the boundary moving towards the plasma, whereas the velocity of its outward motion easily reaches relativistic values.

To increase the generation efficiency in the wavelength range above 3 $\mu$m, we propose using near-critical density targets, which have already been successfully used in various experiments with high-intensity laser pulses \cite{Bin2018, Hilz2018, Ma2019, Zhao2020, Shokita2020, Goethel2022}. Reducing the target density to a critical value leads to an increase in the velocity of the boundary motion without a decrease in the reflection efficiency in terms of the number of reflected photons. A simple estimate shows that if the maximum speed is reached for a time of about $\alpha = 1/10$ of the laser radiation period in the reference frame associated with the plasma boundary, and the Doppler factor is about $\delta = 5$, then for a laser pulse with a wavelength of 0.8 $\mu$m, the energy efficiency of conversion to the range 4 $\mu$m will be about $\alpha/\delta = 0.02$, which for an incident pulse with an energy of about 10~J will give a mid-IR pulse with an energy of 0.2~J. Assuming that the reflected pulse will have a duration on the order of driving pulse i.e. 50~fs we get a power estimate of 4~TW. For a transverse pulse size of about 4 $\mu$m, its intensity will be about $10^{18}$~W/cm$^2$, and the dimensionless relativistic parameter is $a_{\rm mIR} = e\lambda/m_e c^2 \cdot \sqrt{2I/\pi c} = \lambda~[\mathrm{\mu m}]\cdot\sqrt{I[\mathrm{W/cm^2}]/1.38\times10^{18}} \approx 3.4$ (here $c, e, m_e$ are the speed of light, the elementary charge and mass of the electron, respectively, $\lambda, I$ are the wavelength and intensity of the radiation), which is higher than the record values achieved to date. This pulse will be automatically synchronized to sub-femtosecond precision with high harmonics, allowing them to be used together in high-precision pump-probe experiments. At the same time, the high radiation power makes it possible to perform measurements in a large volume of matter, increasing the diffraction signal.

\paragraph{Relativistic electron spring model}

Let us consider the process of radiation generation at the boundary of a plasma irradiated by a short relativistically intense linearly polarized laser pulse. For the sake of clarity, we restrict ourselves to the one-dimensional case along the $x$ axis and the normal incidence of the laser pulse from the region $x < 0$. Since we are also interested in near-critical density plasma, in which internal plasma fields can play a significant role, the most suitable for theoretical analysis is the relativistic electron spring model proposed in \cite{Gonoskov2011} and developed in \cite{Gonoskov2018a}.

The relativistic electron spring model is based on three relatively simple approximations, justified by its comparison with the results of a fully electrodynamic kinetic numerical simulations. First, the ultrarelativistic limit is considered, in which electrons can have a speed equal to either zero or the speed of light. Secondly, it is assumed that the plasma at any moment of time consists of a stationary ion background occupying the region $x>0$, an infinitely thin electron layer at the point $x = x_s(t)$, containing all the electrons squeezed out by the ponderomotive pressure from the region $0 < x < x_s(t)$, and unperturbed electrons in the rest of the space $x > x_s(t)$. Third, it is assumed that the incident field and the field emitted by the electron layer exactly compensate for each other in the region $x > x_s(t)$, which allows us to write the equation of motion of the electron layer \cite{Gonoskov2018a}:
\begin{equation}\label{eq:x_s}
\frac{dx_s}{dt} = \frac{f^2(x_s - t) - \left(n_0 x_s/2\right)^2}{f^2(x_s - t) + \left(n_0 x_s/2\right)^2},
\end{equation}
where relativistic units are used: $f(\xi) = e|\bm{E_\perp}(\xi)|/m_e \omega c$ is the value of the transverse electric field of the incident wave, $n_0 = 4\pi e^2 N_e/m\omega^ 2$ is the unperturbed concentration of electrons in the plasma (plasma is assumed to be initially homogeneous for simplicity), distances are measured in $c/\omega$, and time intervals in $\omega^{-1}$, where $\omega$ is some frequency, which we choose equal to the carrier frequency of incident radiation.

With a known laser pulse shape $f(x-t)$, the equation~(\ref{eq:x_s}) can be integrated numerically. Knowing the position of the layer at each moment of time and determining the currents in it based on the assumptions of the model, it is possible to calculate the reflected radiation \cite{Gonoskov2018a}:
\begin{equation}\label{eq:a_r}
f^r(x_s(t) + t) = -\frac{1-\beta_x}{1+\beta_x}f(x_s(t) - t),
\end{equation}
where $f^r(x+t)$ is the electric field of the reflected pulse, $\beta_x = dx_s/dt$ is the longitudinal velocity of the electron layer.

\begin{figure}
\includegraphics[width=0.45\textwidth]{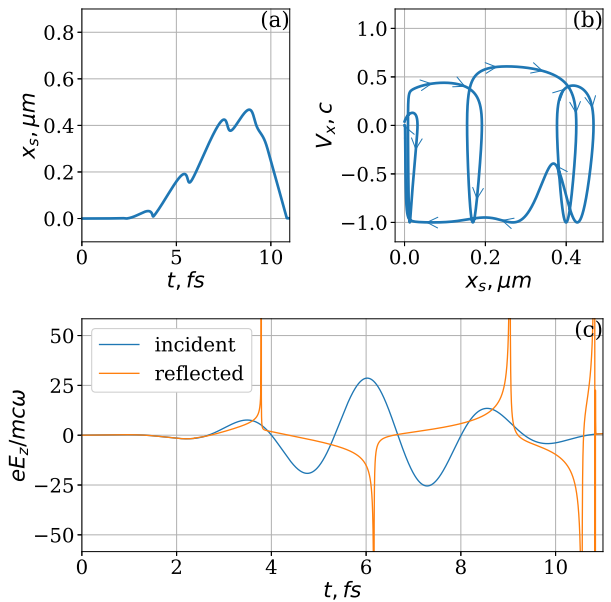}
\caption{\label{fig:model} Solution of the equations of the relativistic electron spring model for a Gaussian pulse with a duration of 3 fs and an amplitude of $a_0=30$ and a homogeneous plasma with a density of $n_0=10$. (a) Trajectory of the electron boundary $x_s(t)$. (b) Trajectory the electron boundary in the phase space $(x, v_x=dx/dt)$. (c) The temporal profiles of the incident (blue) and reflected (orange) pulses.}
\end{figure}

Fig.~\ref{fig:model} shows an example of solving equation (\ref{eq:x_s}) and the corresponding forms of incident and reflected pulses, calculated using the formula (\ref{eq:a_r}). Under the influence of ponderomotive pressure, the electron boundary oscillates, and during the return motion its speed at some moments of time almost reaches the speed of light. At these times, the most efficient generation of high harmonics occurs, which forms sharp peaks with a sub-femtosecond duration in the reflected field. Along with this, however, there is an average motion of the boundary deep into the plasma with a velocity of the order of $0.5-0.6c$. This movement should also lead to an enrichment of the spectrum of the reflected signal in the low-frequency region. It can also be expected that the efficiency of generating a low-frequency signal, as well as its wavelength, will increase with an increase in this speed. Such an increase can be achieved either by increasing the laser pulse amplitude or by decreasing the plasma density. Thus, it can be expected that it would be more optimal to use plasma of near-critical density, which, on the one hand, is still opaque and is able to effectively reflect radiation, and, on the other hand, makes it possible to achieve the maximum boundary velocity.

\paragraph{Particle-In-Cell simulations}
To test the predictions of the theoretical model and study the interaction process in the range of parameters in which the model is not applicable, we performed numerical simulations based on the particle-in-cell (PIC) method. For these purposes, the PICADOR package was used, which implements the fully electrodynamic relativistic PIC method \cite{Surmin2016a}. First, we present the results of one-dimensional calculations, and then discuss the features of the multidimensional problem.

The simulations were carried out in a box 40 $\mu$m in size with a grid step $\Delta x = 2.5$ nm. The time step was $\Delta t = \Delta x/2s \approx 4.2$ as. Plasma at the initial moment of time was set as a layer with a uniform concentration of particles 10 $\mu$m long, the number of macroparticles in the cell was 50. Ions were considered immobile for clarity and because of the short interaction time. The electron temperature at the initial moment of time was set equal to 100 eV, which is much less than the energy of electron motion in the field of the incident wave. The total simulation time was 150 fs, which was sufficient to observe the entire interaction process, as well as the subsequent plasma emission. The parameters of laser radiation corresponded to a Gaussian pulse at a wavelength of 800 nm with an energy of 7.5 J and a duration of 15 fs at FWHM (Full Width at Half Maximum) of the intensity, focused into a spot with a diameter of 5 $\mu$m. For these parameters, the amplitude of the incident pulse is $a_0 = 27.8$.

\begin{figure}
\includegraphics[width=0.45\textwidth]{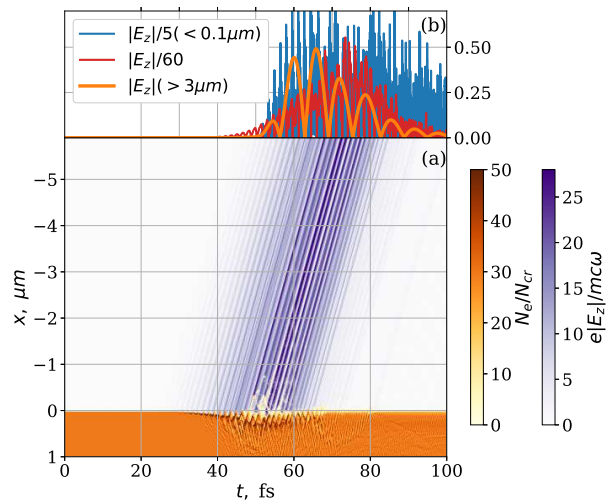}
\caption{\label{fig:1dpic} The result of 1D PIC-simulation for a Gaussian laser pulse with a duration of 15 fs and an amplitude $a_0 = 27.8$ and a homogeneous plasma with a density of $n_0 = 30$. (a) Spatiotemporal dynamics of the electron concentration (orange) and reflected radiation (violet), The concentration is normalized to the critical one for the carrier frequency of the incident laser pulse $N_cr = m\omega^2/4\pi e^2$. (b) Time profile of the reflected pulse without filtering (red) after filtering in the $<0.1$ $\mu$m wavelength range (blue) and after filtering in the $<3$ $\mu$m wavelength range (orange).}
\end{figure}

Figure~\ref{fig:1dpic} shows a typical result of a one-dimensional calculation for the plasma density $n_0 = 30$. It can be seen that, under the action of incident radiation, the plasma boundary undergoes oscillations of significant amplitude with a gradual averaged movement deep into the plasma. Spectral filtering of the reflected pulse shows the presence of high-energy harmonics in it, as well as a low-frequency signal along with them. The low-frequency signal filtered in the wavelength range above 3 $\mu$m reaches the amplitude $a_{\rm mIR} =0.5$ in the given case. Note that the wavelength in this range significantly exceeds the duration of the time intervals during which the boundary motion velocity is positive, so this signal is determined mainly by the envelope of the incident laser pulse, and not by the details of the boundary motion.

\begin{figure}
\includegraphics[width=0.45\textwidth]{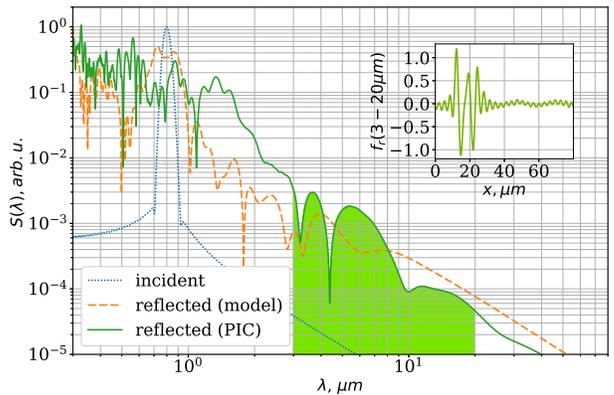}
\caption{\label{fig:cmpr_sp} Comparison of the Fourier spectra for the incident pulse (dashed blue line), the reflected pulse obtained from the relativistic electron spring model (orange dash), and the reflected pulse obtained from the 1D PIC simulation (blue). The inset shows the time profile of the pulse obtained after filtering the reflected signal in 1D PIC simulation in the range 3--20 $\mu$m.}
\end{figure}

To confirm the Doppler mechanism of low-frequency radiation generation, we performed a direct comparison of the spectra of reflected signals obtained within the framework of the relativistic electron spring model and as a result of calculations, shown in Fig.~\ref{fig:cmpr_sp}. Note the good qualitative agreement between these spectra. Some discrepancy is explained by the fact that, strictly speaking, the model of a relativistic electron spring is not applicable in the parameter range under study, since at $n_0/a_0 \sim 1$ signs of relativistic self-induced transparency begin to appear and some of the electrons begin to fly out of the plasma towards the laser radiation, as can be seen on Fig. \ref{fig:1dpic} at time $t\approx 50$ fs. This leads to a violation of the assumption that there are no electrons in the region $x < x_s(t)$. Nevertheless, the proximity of the spectra obtained within the framework of the model and simulation allows us to state that the model correctly captures the main mechanism of low-frequency radiation generation and since the model is purely electrodynamic, this mechanism is the Doppler effect.

One might think that it is the threshold of the induced transparency that determines the optimal plasma density at which the maximum efficiency of low-frequency radiation generation is observed, but our analysis showed that this is not the case.

\begin{figure}
\includegraphics[width=0.45\textwidth]{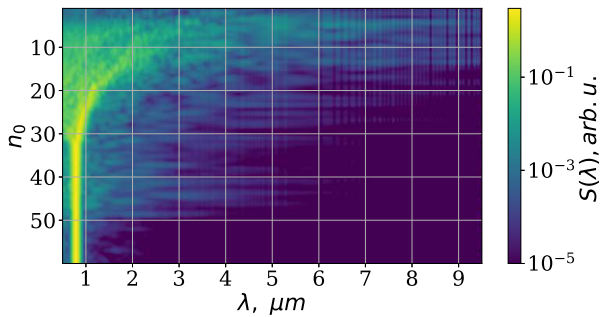}
\caption{\label{fig:sp_vs_n0} Dependence of the Fourier spectrum of the reflected signal obtained in 1D PIC simulation on the plasma density.}
\end{figure}

\begin{figure}
\includegraphics[width=0.45\textwidth]{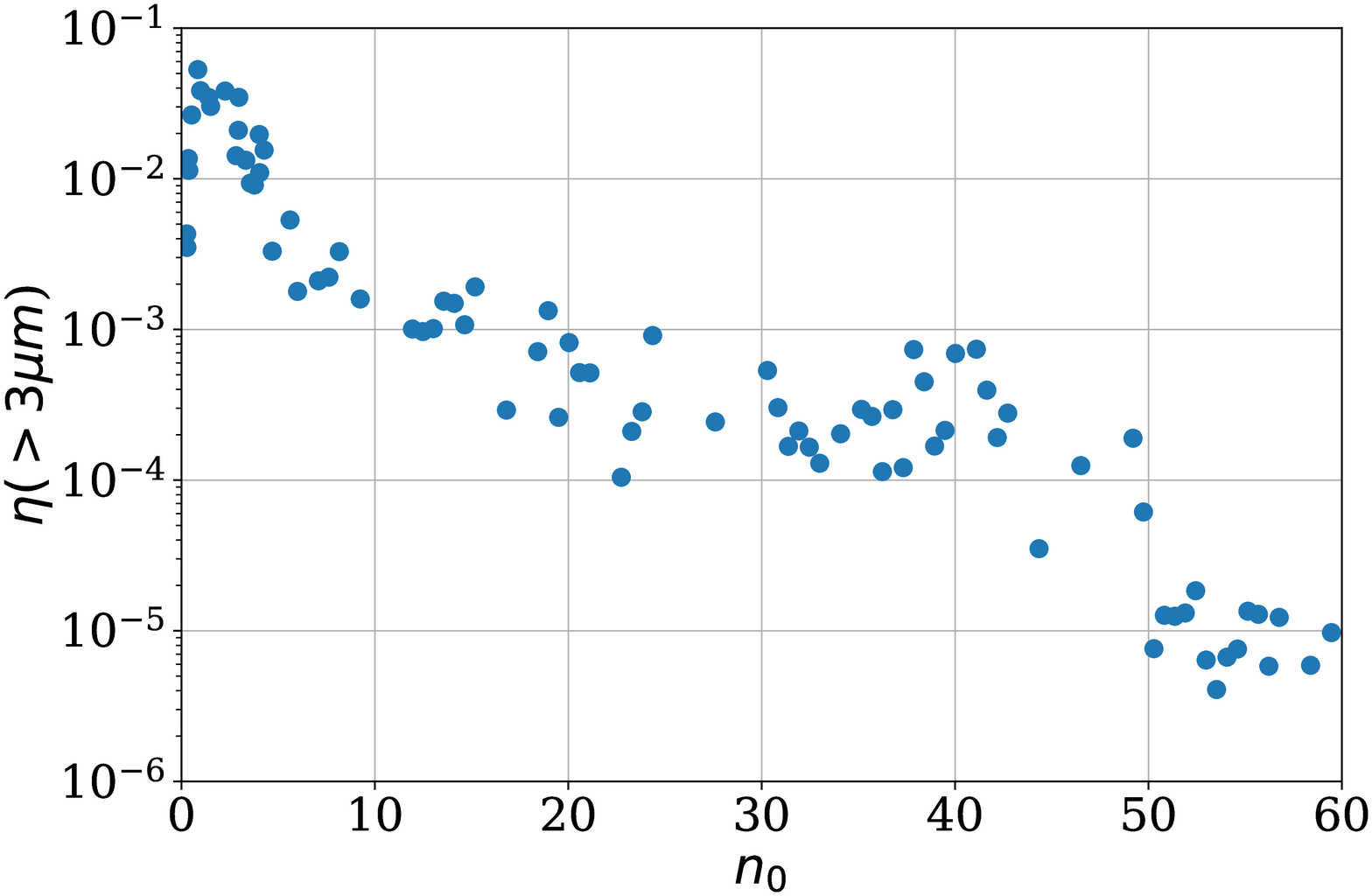}
\caption{\label{fig:efficiency} Dependence of the energy efficiency of the conversion of the incident laser radiation into the reflected signal in the range of $>3$ $\mu$m.}
\end{figure}

Fig.~\ref{fig:sp_vs_n0} shows the dependence of the reflected signal spectrum on the plasma density. Indeed, for $n_0 < a_0$ a strong modification of this spectrum is observed. An increase in the spectral width, an increase in the maximal generated wavelength, and a shift of the spectrum maximum to the long wavelength region are observed. However, note that the position of spectrum maximum does not exceed 3 $\mu$m even for $n_0 \sim 1$, and thus does not enter the mid-IR region. Despite these changes, the total radiation power in the long-wavelength region does not decrease with the decrease in density in the range $1 < n_0 < a_0$. This can be seen from the radiation energy integrated over the range $\lambda > 3$ $\mu$m, whose dependence on the plasma density is shown in Fig. \ref{fig:efficiency}.

This behavior can be explained by the fact that, as was noted previously \cite{Eremin2010, Siminos2012, Siminos2017, Mikheitsev2020}, induced transparency at a sharp plasma boundary manifests itself in the appearance of electron beams breaking off towards the incident laser radiation. This is accompanied by a deeper penetration of the radiation into the plasma. However, if the plasma density is not too low, this process inevitably stops, and later the laser radiation is completely reflected. Even after the disruption, the reflective electron layer is not completely destroyed and continues to reflect radiation. Therefore, generation of low-frequency radiation is also observed in this case, and its efficiency does not decrease.

Note, however, that for a sufficiently low $n_0 < 1$, the radiation begins to penetrate into the plasma continuously, total reflection does not occur, and the efficiency of the Doppler generation mechanism here becomes low. It turns out to be optimal to use a plasma with a density $n_0 \approx 1-2$, at which the conversion of up to 5~\% of the incident energy into radiation energy with a wavelength greater than 3 $\mu$m is achieved. This result agrees well with the rough estimate we gave at the beginning.

Figure~\ref{fig:cmpr_sp} also shows the shape of the filtered mid-IR pulse. This pulse contains several field oscillations, the number of which, however, is less than in the incident pulse. Thus, after filtering, it is possible to obtain an extremely short pulse with relativistic amplitude.

\paragraph{2D PIC simulations}

One-dimensional modeling makes it possible, on the one hand, to highlight the most important physical mechanisms responsible for the observed effects, and, on the other hand, to carry out a relatively fast multiparameter study. However, in reality, it is difficult to create conditions close to one-dimensional, in particular, because of the need to use relatively sharp focusing to achieve the desired intensities, and also because of the possible development of transverse instabilities. To confirm that these features do not fundamentally affect the result obtained in the one-dimensional analysis, we also carried out a two-dimensional calculation using the same fully electrodynamic kinetic PIC method implemented in the PICADOR package.

In this calculation, the box size was $140\times 60$~$\mu$m with a computational grid of $7000\times 3000$ points. The time step was $\Delta t = \Delta x/2s \approx 33$ as. The plasma at the initial moment of time consists of a homogeneous layer with a density of $n_0=40$ and a thickness of 10 $\mu$m and a pre-plasma layer in which the density linearly increased from $n_0=0$ to $n_0=40$ at a distance of 2 $\mu$m. The preplasma layer mimics the influence of a possible prepulse in the experiment. The immobile ions were considered, which was justified by the short interaction time and relatively low density of the target. The number of macroparticles in the cell was 10. The electron temperature at the initial moment of time was set equal to 100~eV. The total simulation time was 1.5 ps. The incident pulse had a wavelength of 800 nm, a Gaussian shape with a duration of 20 fs at FWHM, and was focused on the surface of the plasma layer in a spot with a diameter of 6~$\mu$m at FWHM. The pulse energy was assumed to be 10 J, which corresponds to an amplitude at the focus of $a_0 = 33.5$ and an intensity of $4.8\times10^{21}$ W/cm$^2$.

\begin{figure*}
\includegraphics[width=\textwidth]{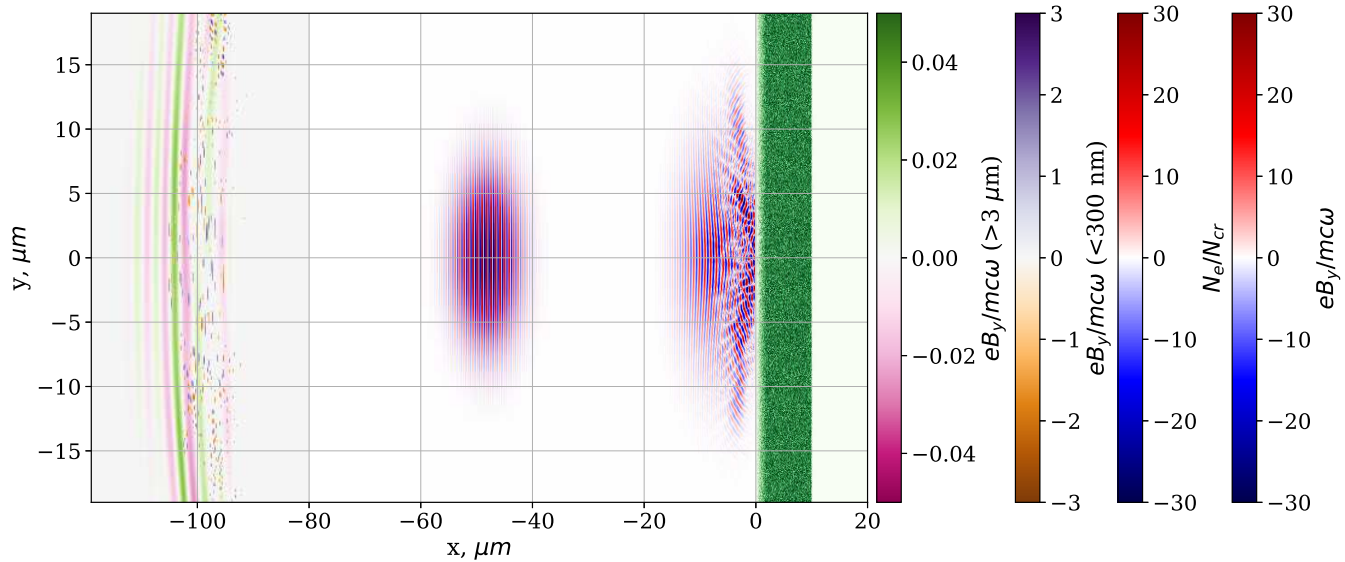}
\caption{\label{fig:2dpic} The result of a 2D PIC simulation of the interaction of a Gaussian pulse with a duration of 20 fs, a width of 6 $\mu$m and an amplitude of $a_0 = 33.5$ with a plasma with a density of $n_0=40$, which has a preplasma layer with a thickness of 2 $\mu$m with a linear density gradient. In the center is shown the laser pulse at the beginning of the calculation, and in the right is shown the transverse fields and the electron density at the time $t=200$ fs after the start of the calculation. On the left, the low-frequency (in the wavelength range $>3$ $\mu$m, pink-green) and high-frequency (in the wavelength range $<300$ nm, orange-violet) parts of the reflected pulse at time $t=550$ fs after the start of the calculation are shown.}
\end{figure*}

The calculation results are shown in Fig.~\ref{fig:2dpic}. Note that, as in the one-dimensional case, efficient generation of both high-frequency bursts and low-frequency radiation containing several field oscillations is observed. Due to diffraction, however, the amplitude of the low-frequency signal turns out to be lower than in the one-dimensional case, and, in addition, its rather strong divergence is observed, caused by the relatively small focusing radius of the laser pulse and, as a consequence, by the small size of the emitter compared to the generated wavelength. Nevertheless, the main effect is retained when considering the two-dimensional case, which allows us to hope for the possibility of successful demonstration in the experiment as well.

\paragraph{Conclusions}

In conclusion, we have proposed a new scheme for the simultaneous generation of X-ray and mid-IR pulses internally synchronized with subfemtosecond accuracy. The scheme is based on the well-known method of Doppler frequency conversion in the interaction of relativistically intense laser pulses with the surface of overdense plasma targets. Due to the use of near-critical density targets, the efficiency of mid-infrared pulse generation can be greatly increased and can reach several percent in energy. The generated mid-infrared pulses have only a few optical cycles and can reach relativistic amplitudes exceeding the record values achieved to date by alternative methods.

Simultaneous generation of high-power synchronized X-ray and mid-IR pulses can be required for pump-probe experiments, for example, laser-induced electron diffraction with increased accuracy.

\begin{acknowledgments}
The research was supported by the Ministry of Science and Higher Education of the Russian Federation, state assignment for the Lobachevsky University of Nizhny Novgorod, project 0729-2020-0035, and state assignment for the Institute of Applied Physics RAS, project 0030-2021-0012. The simulations were performed on resources provided by the Joint Supercomputer Center of the Russian Academy of Sciences.
\end{acknowledgments}

\bibliography{mid-ir-ncd}

\end{document}